\RequirePackage{fix-cm}
\documentclass[aps,pre,preprint]{revtex4-1}          
\usepackage{graphicx}
\usepackage{mathptmx}      
\usepackage{latexsym}
\usepackage{amsmath}%
\usepackage{amsfonts}%
\usepackage{amssymb}%
\usepackage{times}
\usepackage{natbib}

\newlength{\fw}
\begin{document}

\title{Mechanism governing separation in microfluidic pinched flow fractionation devices}

\author{Sumedh R. Risbud}
\affiliation{Chemical and Biomolecular Engineering, Johns Hopkins University}
\author{German Drazer}
\affiliation{Mechanical and Aerospace Engineering, Rutgers, the State University of New Jersey}

\begin{abstract}
We present a computational investigation of the mechanism governing {\em size-based} particle 
separation in microfluidic pinched flow fractionation. We study the behavior of particles 
moving through a pinching gap (i.e., a constriction in the aperture of a channel) in the Stokes 
regime (negligible fluid and particle inertia) as a function of particle size. The constriction 
aperture is created by a plane 
wall and spherical obstacle, and emulates the pinching segment in pinched flow 
fractionation devices. The simulation results show that the distance of closest approach 
between the particle and obstacle surfaces (along a trajectory) decreases with increasing 
particle size. We then use the distance of closest approach to investigate the effect of 
short-range repulsive non-hydrodynamic interactions (e.g., solid-solid contact due to 
surface roughness, electrostatic or steric repulsion, etc.). We define a {\em critical 
trajectory} as the one in which the minimum particle-obstacle separation is equal to the 
range of the non-hydrodynamic interactions. The results further show that the initial 
offset of the critical trajectory (defined as the critical offset) increases with particle 
size. We interpret the variation of the critical offset with particle size as the basis 
for size-based microfluidic separation in pinched flow fractionation. We also compare the 
effect of different driving fields on the particle trajectories; we simulate a constant 
force driving the particles in a quiescent fluid as well as a freely suspended particles 
in a pressure-driven flow. We observe that the particles driven by a constant force 
approach closer to the obstacle than those suspended in a flow (for the same initial offset). 
On the other hand, the increment in the critical offset (as a function of particle size) 
is larger in the pressure-driven case than in the force-driven case. Thus, pressure-driven 
particle separation using pinched flow fractionation would prove more effective than its 
force-driven counterpart (e.g., particles settling under gravity through a pinching gap).
\end{abstract}

\maketitle

\section{Introduction}\label{sec:introLB2}
A particularly attractive microfluidic separation technique would be able to exploit the 
differences in the interactions of the particles with the geometric features embedded within 
the channels of a micro-device, without the need of an external field. A few examples of such 
fluidic-only separation methods include: size-exclusion, entropic trapping \citep{han2000}, 
deterministic lateral displacement using solid obstacles \citep{huang2004}, and pinched flow 
fractionation (PFF, \citet{yamada2004}). PFF is a relatively simple method, in which species 
entering a channel constriction then exit into a sudden expansion at different positions across 
the channel. As 
a result of its simplicity and promise, numerous variants of PFF have emerged within the last 
decade \citep{heonalee2011,larsen2008,maenaka2008,morijiri2011,vig2008}.

The separation of flow streamlines coming out of the constriction was originally suggested as the basis 
of PFF \citep{yamada2004}. This explanation, however, does not take into account particle-wall hydrodynamic 
interactions. Two factors indicate the importance of considering hydrodynamic interactions to determine 
the particle trajectories through a constriction. First, the particles are similar in size to 
the width of the constriction and, therefore, cannot be considered as tracer particles advected by the flow 
\citep{mortensen2007,shardt2012,ashley2013}. Second, as the particles move through the constriction, the 
surface-to-surface separation between the particles and the channel wall tend to becomes much smaller that 
the size of the particles, and lubrication forces can play a significant role. Studying the motion of a 
particle through a constriction is also relevant in the context of certain particle focusing methods 
\citep{faivre2006,xuan2010} and micro-models of porous media employed in particle deposition studies 
\citep{wyss2006,mustin2010}.

\begin{figure}
\begin{center}
\includegraphics[width=1.25\fw]{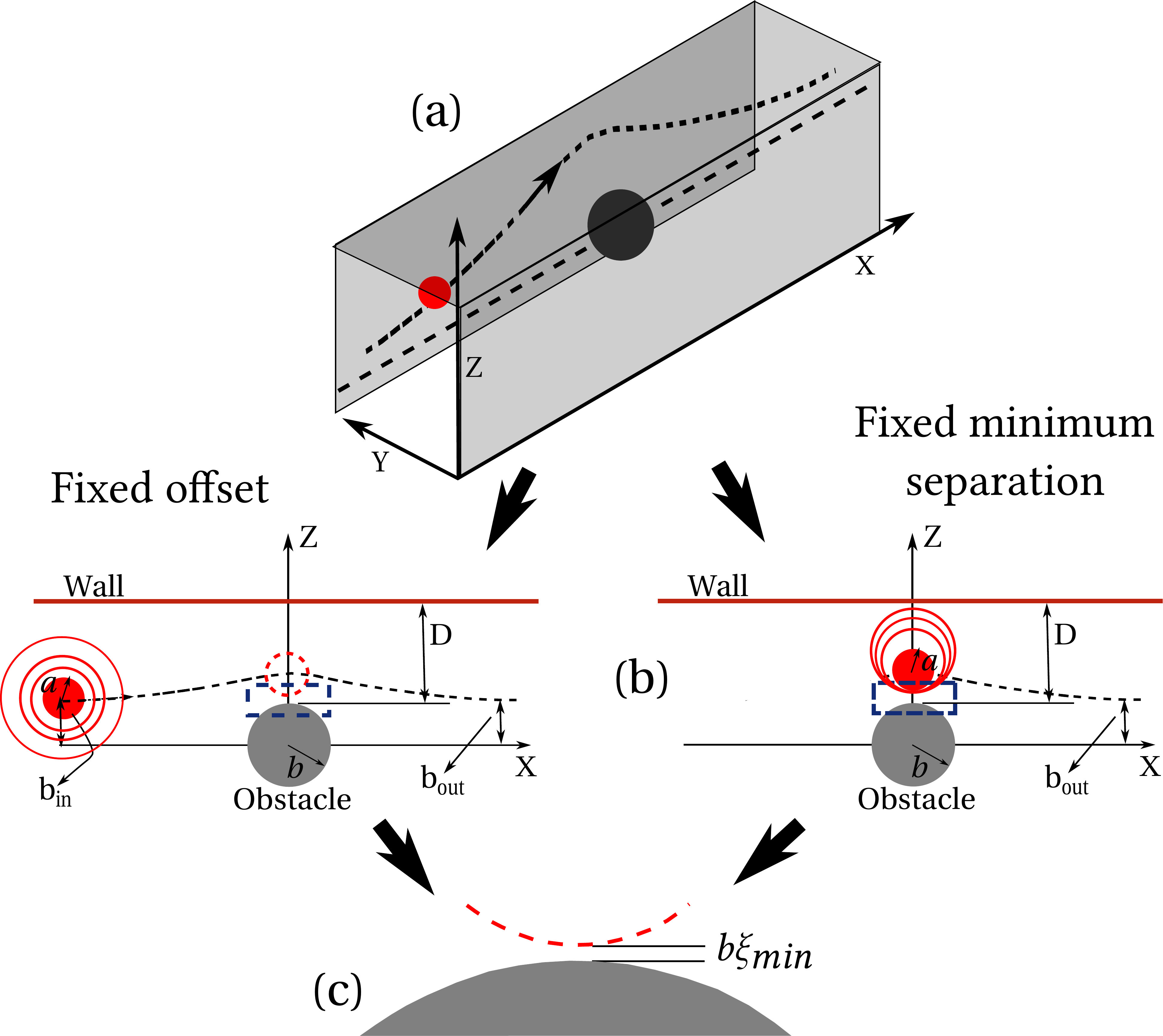}
\caption{Simulation box and relevant length scales. (a) The simulation box with the coordinate system. 
Two parallel walls perpendicular to the $Y$-axis form a channel in all simulations. (b) This is an 
enlarged view of the relevant part of the simulation box. (left) In {\em fixed offset} simulations, the initial 
position of all particle centers is the same (as depicted by the concentric circles). The initial and 
the final offsets ($b_{in},~b_{out}$) are indicated, along with the particle and obstacle radii and the 
size of the constriction aperture $D$. The dashed circle shows the position of the closest approach 
along the particle trajectory, in absence of significant inertia. (right) In {\em fixed-$\xi_{min}$} simulations, 
the particles begin their motion just before the apex of their trajectories, and attain the same minimum 
surface-to-surface separation (as depicted by the internally tangential circles). The radii of the particle 
and the obstacle are $a$ and $b$, respectively. (c) Close up of the region enclosed by a dashed box in (b) 
highlighting the minimum separation between particle and obstacle surfaces.}
\label{fig:systemGeometryLB2}
\end{center}
\end{figure}

We have recently studied the relationship between the initial offset in a particle trajectory ($b_{in}$ 
in figure \ref{fig:systemGeometryLB2}(a)) and the minimum surface-to-surface separation between the 
particle and the obstacle along the trajectory. We have shown that even a moderate initial offset in a 
particle trajectory (say, comparable to the particle radius) leads to surface-to-surface separations that 
are significantly small, e.g., $O(100~nm)$ between a particle and an obstacle of micrometer size 
\citep{risbud2013}. The occurence of such small surface-to-surface separations highlights the importance 
of short-range non-hydrodynamic interactions, and suggests the minimum separation as the relevant 
length-scale to compare with their range. Such interactions can be modeled using a hard-wall repulsion, 
which leads to the definition of a {\em critical offset}, i.e., the smallest initial offset that results in 
a fore-aft symmetric trajectory \citep{balvin2009,frechette2009,risbud2013}. 
We have also shown that the relationship between the critical offset and the range of repulsive 
interactions is the same as that between the initial offset and the minimum surface-to-surface separation 
\citep{frechette2009, risbud2013}. In previous lattice Boltzmann simulations (\citet{risbud2013a}, 
henceforth {\em paper I}) we studied the trajectory followed by a spherical particle 
as it passes through a constriction created by a spherical obstacle of {\em the same size} and a plane 
wall. For the same initial offset, we showed that a particle reaches closer to the obstacle as the 
constriction aperture decreases. Therefore, the critical offset increases with decresing 
constriction aperture. 

In this article, we investigate the motion of particles of {\em unequal sizes} as they move through a constriction.
We show that irreversible particle-wall interactions could lead to {\em size-based separation} at low Reynolds 
number. Particularly, we show that particles of different sizes exhibit different extents of lateral displacement 
as they move through a constriction. We present the results of lattice Boltzmann simulations for various 
particle-obstacle aspect ratios and sizes of the constriction aperture in the Stokes regime 
(negligible particle and fluid inertia). We also investigate the dependence of the critical offset on the 
driving field; we simulate the trajectories resulting from a constant force driving the particles 
in a quiescent fluid as well as those in which a suspended particle moves with a pressure-driven flow. 
Based on the hard-wall model, we show that larger particles exhibit a larger 
critical offset in the presence of a non-hydrodynamic repulsion of fixed range. Therefore, 
size-based separation by employing the motion of particles through a constriction can be qualitatively explained 
through this study. Further, we show that decreasing the constriction aperture also increases the critical 
offset, with the increment being larger for the case of a pressure-driven flow carrying the particles. 

The article is organized as follows: in \S\ref{sec:sysDefLB2}, we introduce the system under investigation, 
as well as the nature of the simulations carried out. We discuss the hard-core model for short-range repulsive 
non-hydrodynamic interactions in \S\ref{sec:hardCoreModelLB2}. The results of the simulations are presented in 
\S\ref{subsec:setISimResultsLB2} and \ref{subsec:setIISimResultsLB2} and the discussion relevant to size-based 
separation in \S\ref{subsec:sizeBasedSep}.

\section{The system}\label{sec:sysDefLB2}
Figure \ref{fig:systemGeometryLB2} depicts the system studied in this work. We consider a suspended spherical particle of 
radius $a$ (diameter $d$) negotiating a fixed spherical obstacle of radius $b$ along the positive $x$-axis. 
We use the lattice Boltzmann method ({\em susp3d}) \citep{ladd1994a,ladd1994b,nguyen2002}.
The simulation box is outlined in figure \ref{fig:systemGeometryLB2}(a), and has 
dimensions $x\times y\times z\equiv240\times60\times140$ lattice units. The two walls perpendicular to the $y$-axis 
form a channel (henceforth, `side-walls' or channel-walls). The plane of motion is the mid-plane of the channel, 
parallel to the $xz$-plane. The particle motion is planar, confined to the mid-plane due to the symmetry of the problem. 
The particle trajectory passes through a constriction of minimum aperture $D$, created by the plane wall perpendicular 
to the positive $z$-axis (henceforth, the `top' wall) between itself and the obstacle surface (figure 
\ref{fig:systemGeometryLB2}(b)). The initial (upstream) and final (downstream) offsets in the particle trajectory are 
denoted by $b_{in}$ and $b_{out}$, respectively. Since we are investigating particle trajectories in the Stokes regime 
(i.e., negligible fluid as well as particle inertia), the trajectories are fore-aft symmetric and the two offsets are 
equal ($b_{in}=b_{out}$). The minimum surface-to-surface separation between a particle and the obstacle is $b\xi_{min}$ 
as shown in figure \ref{fig:systemGeometryLB2}(c). The no-slip boundary condition is imposed on all solid boundaries, 
while a periodic boundary condition is imposed on the faces of the simulation box perpendicular to the $x$-axis. A 
periodic boundary condition is also imposed on the faces perpendicular to the $z$-axis, when the wall creating the 
constriction is absent. We vary the constriction aperture by translating the obstacle center along 
$z$-axis towards the top-wall.

We have performed two sets of simulations (figure \ref{fig:systemGeometryLB2}(b)): first, we investigate 
trajectories that have the same initial offset $b_{in}=30$ lattice units, with an obstacle of radius $b=10$ 
lattice units (henceforth, {\em fixed offset} simulations), and second, we study trajectories that attain 
the same minimum separation between particle and obstacle surfaces ($1$ lattice unit), with an obstacle of 
radius $b=20$ lattice units (henceforth, {\em fixed-$\xi_{min}$} simulations). We use a larger obstacle 
in the fixed-$\xi_{min}$ simulations to achieve a smaller minimum separation ($1$ lattice unit $\equiv$ 
$\xi_{min}=5\times10^{-2}$). The fixed offset simulations exhibit different minimum separations, whereas 
the fixed-$\xi_{min}$ simulations exhibit different final offsets. Both cases probe the dependence 
of the respective variables on the constriction aperture and particle size. We investigate the 
effect of two driving fields: a constant force driving the particles in a quiescent fluid, and freely 
suspended particles moving with a pressure-driven flow (under a constant pressure-drop). In both sets 
of simulations, the obstacle radius serves as the characteristic length scale of the problem. As 
mentioned earlier, the magnitudes of particle as well as fluid inertia are negligible in all simulations
($\text{Re}\sim O(10^{-2})-O(10^{-3})$ and $\text{St}\sim O(10^{-3})-O(10^{-4})$).

The particle-to-obstacle aspect ratio $\alpha=a/b$ and the dimensionless constriction aperture $\Delta=D/b$ 
form the parameter space relevant to this study. We use four different particle sizes, $a = 5,~10,~15,~20$ 
lattice units, yielding the particle-to-obstacle aspect ratios $\alpha = \tfrac{1}{2},~1,~\tfrac{3}{2},~2$ 
for fixed offset simulations, and $\alpha = \tfrac{1}{4},~\tfrac{1}{2},~\tfrac{3}{4},~1$ for fixed-$\xi_{min}$ 
simulations. Note that the values of $\alpha$ are different for the two sets of simulations due to different 
obstacle radii.

In the fixed offset simulations, we investigate the constriction apertures $D=60$ and $D=50$ lattice units
($\Delta=6$ decreasing to $\Delta=5$), and also investigate the trajectories in the absence of the top wall 
(the latter corresponding to the absence of a constriction). For the fixed-$\xi_{min}$ simulations, the sizes 
of the constriction apertures are $D=98$, $78$, $58$, and $45$ lattice units ($\Delta = 4.9,~3.9,~2.9,~2.25$). 
The initial and final offsets are noted at $x=-100$ and $x=100$ lattice units, respectively. For each 
fixed-$\xi_{min}$ simulation, the initial $z$-coordinate of the particle center is calculated such that 
approximately the same minimum surface-to-surface separation is attained by particles of all sizes. 

As discussed in \S\ref{sec:introLB2}, the minimum separation attained along a particle trajectory ($\xi_{min}$ 
in figure \ref{fig:systemGeometryLB2}(c)) serves as the length scale that should be compared with the 
range of short-range non-hydrodynamic interactions to estimate their effect on the trajectory. Thus, in what 
follows we present the results obtained for the minimum separation $\xi_{min}$ as a function of the 
particle-obstacle aspect ratio and constriction aperture.

\section{The minimum separation and the model for non-hydrodynamic interactions}\label{sec:hardCoreModelLB2}
\begin{figure}
\begin{center}
\includegraphics[width=1.25\fw]{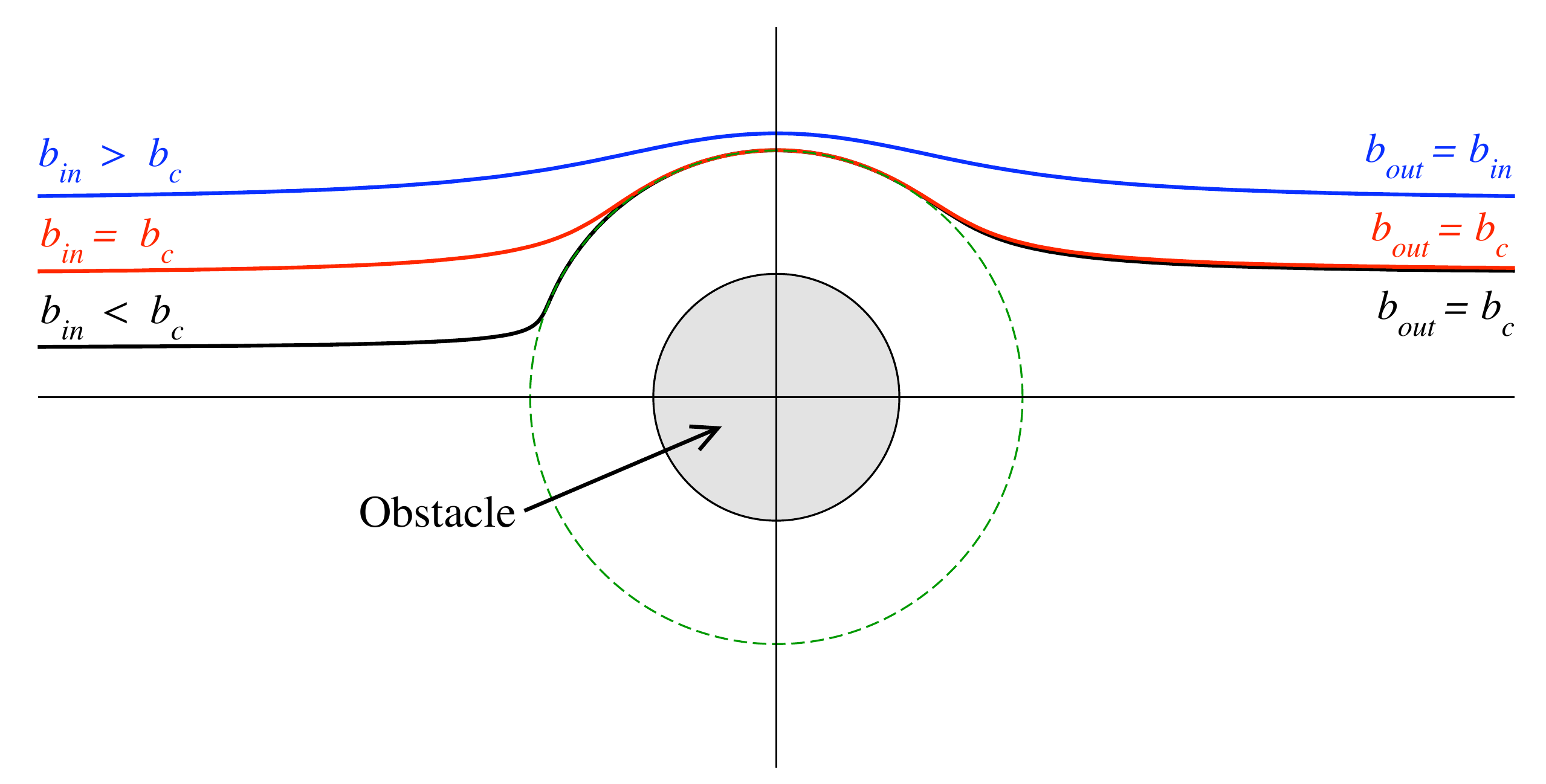}
\caption{Three types of particle trajectories in the presence of non-hydrodynamic interactions in the Stokes regime.
Reproduced from \citet{balvin2009}.}
\label{fig:threeTypesOfCollisionsNoInertiaLB2}
\end{center}
\end{figure}

Following paper I, we model the short-range repulsive (irreversible) interactions between a 
particle-obstacle pair (such as solid-solid contact due to surface roughness), as a hard-wall 
potential creating a hard-core shell of effective range $\epsilon$, such that, any surface-to-surface separation 
between the particle and the obstacle less than $\epsilon$ (i.e., $\xi_{min}<\epsilon$) is 
not attainable. We have established that this model leads to the definition of a critical 
offset $b_c$ that depends on the range of the non-hydrodynamic interactions 
\citep{frechette2009,risbud2013}. The existence of the critical offset can be rationalized as 
follows: the presence of non-hydrodynamic interactions with a dimensionless range $\epsilon$ 
around the obstacle prevents the particle surface to approach the obstacle surface closer than 
$\epsilon$. Then, the particle trajectories corresponding to $\xi_{min}>\epsilon$ are unaffected 
by the presence of the non-hydrodynamic interactions (see the top-most trajectory in figure 
\ref{fig:threeTypesOfCollisionsNoInertiaLB2}). The trajectory that corresponds to a minimum 
surface-to-surface separation $\xi_{min}=\epsilon$ serves as the {\em critical trajectory} (see 
the middle trajectory in figure \ref{fig:threeTypesOfCollisionsNoInertiaLB2}). Finally, the 
trajectories that would have approached a minimum surface-to-surface separation 
$\xi_{min}<\epsilon$ in the absence of the non-hydrodynamic interactions, are forced to 
circumnavigate the obstacle while maintaining a constant separation equal to $\epsilon$ due to 
the hard-wall repulsion. These trajectories collapse onto the critical trajectory downstream of 
the obstacle (see the bottom-most trajectory in figure \ref{fig:threeTypesOfCollisionsNoInertiaLB2}). 
Alternatively, the critical offset $b_c$ can be defined as the {\em smallest initial offset that 
results in a symmetric particle trajectory around the obstacle}. Further, since the minimum 
separation attained by a particle moving along the critical trajectory is exactly $\xi_{min}=\epsilon$, 
from the above definition, the relationship between $b_c$ and $\epsilon$ is the same as that between 
$b_{in}$ and $\xi_{min}$ \citep{frechette2009,bowman2012,devendra2012}.

\section{Results and discussion}\label{sec:resultsLB2}
\subsection{Fixed offset simulations}\label{subsec:setISimResultsLB2}
\begin{figure}
\begin{center}
\includegraphics[width=1.25\fw]{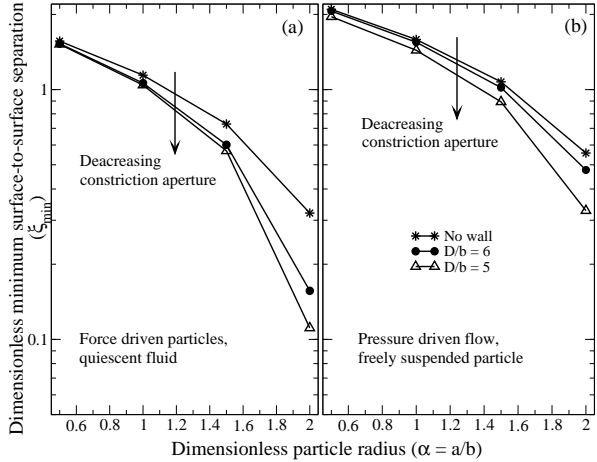}
\caption{The dimensionless minimum separation $\xi_{min}$ versus the dimensionless particle radius $\alpha$ 
for different driving fields: (a)A constant force acting on the particle in a quiescent fluid, (b) a 
constant pressure drop acting on the fluid with freely suspended particle}
\label{fig:ximinVsAlpha_NoInertiaAndD60}
\end{center}
\end{figure}
Figure \ref{fig:ximinVsAlpha_NoInertiaAndD60} shows $\xi_{min}$ as a function of the particle size, for the 
two different driving fields. In 
figure \ref{fig:ximinVsAlpha_NoInertiaAndD60}(a), we plot the case when the particle is driven by a constant 
force in a quiescent fluid. Figure \ref{fig:ximinVsAlpha_NoInertiaAndD60}(b) on the other hand, represents 
the pressure-driven case. In both 
cases we observe that, $\xi_{min}$ decreases with increasing particle size for a constant constriction aperture 
and also with decreasing constriction aperture for a fixed particle size. The latter observation extends 
the results presented in paper I for a particle of the same size as the obstacle. Note that, we have shown via 
theoretical calculations that the minimum separation as a function of the particle size reverses its trend 
for asymptotically small separations in the absence of a constriction ($\xi_{min}\ll\text{O}(10^{-7})$), so 
that smaller particles reach closer to the obstacle \citep{risbud2013}. However, these separations are not 
relevant in practice, since they are much smaller than the range of typical irreversible forces for micrometer 
size particles.
Comparing figures \ref{fig:ximinVsAlpha_NoInertiaAndD60}(a) and (b), we note that a given particle reaches 
closer to the obstacle when moving due to a constant force in a quiescent fluid, than a pressure-driven flow. 
This is consistent with our analysis based on the hydrodynamic mobility of a sphere 
around an obstacle, in the absence of the wall creating the constriction \citep{risbud2013}.
In summary we conclude that {\em the particles reach closer to the obstacle upon 
decreasing the constriction aperture, or increasing the particle size.} 

Let us now discuss the implications of the hard-wall potential model.
As mentioned earlier in \S\ref{sec:sysDefLB2}, all trajectories in this set have the same initial offset 
$b_{in}$, and the obstacle radius $b$ is constant ($b=10$ lattice units).  
Let the corresponding critical offset be $b_{c0}< b_{in}$ for a 
given a given particle radius and constriction aperture $D=D_0$. We first discuss the effect of 
decreasing the constriction aperture keeping the particle radius 
constant (i.e., following the direction of the downward arrow in figure \ref{fig:ximinVsAlpha_NoInertiaAndD60}). 
Since we have observed that $\xi_{min}$ decreases with the constriction aperture, for a 
sufficiently small aperture $D^\prime<D_0$, the minimum separation would, in principle, reach the range of 
the non-hydrodynamic interactions ($\xi_{min}=\epsilon_0$). Correspondingly, the critical offset would
increase from $b_{c0}<b_{in}$ to $b_c^\prime=b_{in}$. 
In other words, 
{\em for a given particle radius, the corresponding critical offset increases with decreasing constriction 
aperture}. 
We note that this result is in qualitative agreement with previous experiments \citep{luo2011}, and 
extends the results obtained for $\alpha=1$ (paper I) to $\alpha\ne 1$. 

The above inference was drawn using a fixed $\alpha$ and varying the constriction aperture $\Delta$. 
An analogous argument can be made using the particle radius $a$ as a variable, wherein the minimum 
separation is observed to decrease with increasing particle radius. Therefore,
{\em a decrease in $\xi_{min}$ due to increasing the particle radius 
for a given constriction aperture, results in an increase in the critical offset $b_c$ in the presence of 
non-hydrodynamic interactions}. However, in this case, we need to assume that the 
non-hydrodynamic interactions have a fixed dimensionless range $\epsilon=\epsilon_0$, set by the obstacle 
and independent of particle size.

Therefore, in general, a decrease in the minimum separation ($\xi_{min}\downarrow$) in the absence of 
non-hydrodynamic interactions, can be translated as an increase in the critical offset ($b_c\uparrow$) in their 
presence. In the following section, we indeed show this to be the case by direct verification using 
the fixed-$\xi_{min}$ simulations.

\subsection{Fixed-$\xi_{min}$ simulations}\label{subsec:setIISimResultsLB2}
\begin{figure}
\begin{center}
\includegraphics[width=1.35\fw]{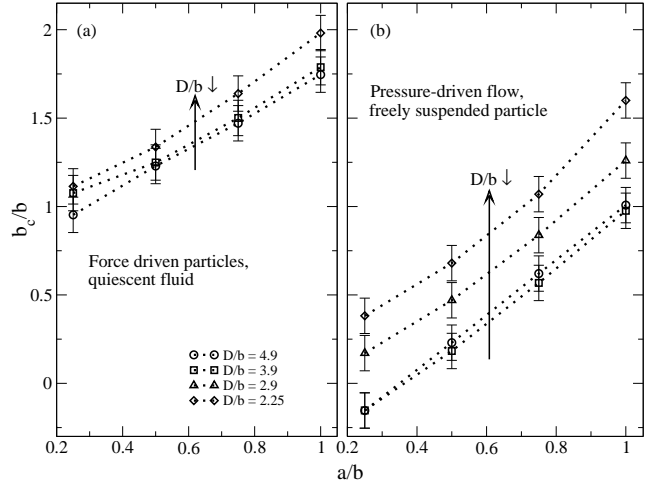}
\caption{The critical offset $b_c$ as a function of the particle radius. The critical offset is evaluated at 
$\epsilon=0.05$ by linear interpolation between two simulated trajectories for each particle size, leading to the 
error-bars depicted above. (a) A constant force drives the particles past the obstacle in a quiescent fluid, (b) a 
constant pressure drop drives the fluid carrying the particle past the obstacle.}
\label{fig:bcVsaLB2}
\end{center}
\end{figure}

The objective of this set of simulations is to investigate the dependence of $b_{out}$ on the parameters of 
the problem when the minimum separation is kept constant. Specifically, we perform simulations that correspond 
to a minimum separation of $1$ lattice unit, i.e., $\epsilon=1/20=0.05$ (for $b=20$ lattice units). 
We know that, the functional relationship between $b_{out}~(=b_{in}\text{ for negligible inertia})$ and $\xi_{min}$ 
is the same as that between $b_c$ and $\epsilon$ \citep{risbud2013}. Therefore, the final offsets $b_{out}$ 
corresponding to the various trajectories obtained in these simulations are the values of the critical offset $b_c$ 
corresponding to $\epsilon=0.05$.
\begin{figure}
\begin{center}
\includegraphics[width=1.35\fw]{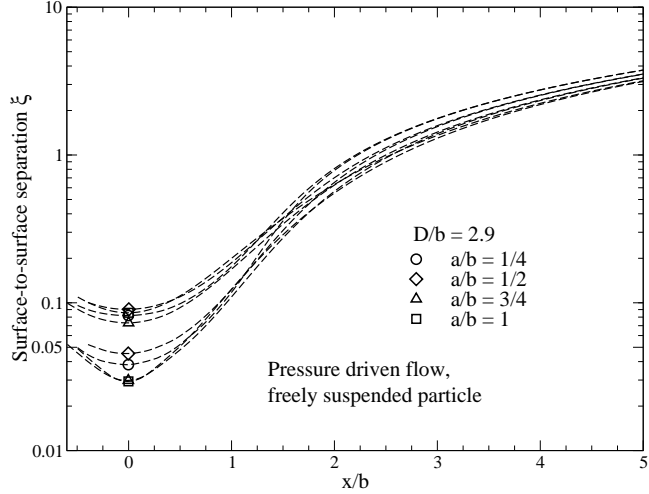}
\caption{The dimensionless surface-to-surface separation $\xi$ between the particle and the obstacle as a function 
of the dimensionless $x$-coordinate. As mentioned in the text, two simulations for each particle size are carried 
out for the purposes of linear interpolation, such that both attain minimum separation close to 
$\xi_{min}=0.05$ (one less than $0.05$, and the other greater than $0.05$). Four different line-styles correspond to the 
four particle sizes used in the simulations. The two groups of trajectories (one corresponding to $\xi_{min}>0.05$ 
and the other corresponding to $\xi_{min}<0.05$) can be clearly 
seen in the figure. This particular plot depicts the case when $b=20$ lattice units, $D/b=2.9$ and the particles 
are carried around the obstacle by pressure-driven flow.}
\label{fig:xiVsXFlowD3LB2}
\end{center}
\end{figure}
However, as shown in figure \ref{fig:xiVsXFlowD3LB2}, our simulated trajectories attain the minimum separation 
of $0.05$ only approximately. Therefore, for each particle size, we compute the final offset corresponding to 
$\xi_{min}=0.05$ by linear interpolation between two independent simulations for which $\xi_{min}$ is greater 
and smaller than 0.05, leading to the error-bars shown in figure \ref{fig:bcVsaLB2}. It is evident from figure 
\ref{fig:bcVsaLB2} that the critical offset increases with the size of the particle (an inference already stated in 
\S\ref{subsec:setISimResultsLB2}), qualitatively corroborating earlier experimental observations \citep{luo2011}. 

We again compare the effect of the field driving the particle past the obstacle. Figure \ref{fig:bcVsaLB2}(a) 
corresponds to a particle driven by a constant force, whereas figure \ref{fig:bcVsaLB2}(b) shows the 
results for pressure-driven flow. We observe 
that the driving field significantly affects the critical offset: in general, higher critical offsets are 
observed for the case of a constant force driving the particle in a quiescent fluid. This is consistent 
with the argument presented in \S\ref{subsec:setISimResultsLB2} that a given particle reaches 
closer to the obstacle when driven by a constant force compared to a pressure-driven flow. 

Figure \ref{fig:bcVsaLB2} also shows that decreasing the constriction aperture affects the case of fluid 
flow more than the case of a constant force driving the particles. Further, in previous experimental findings we 
have observed that the increment in the critical offset itself increases with particle size \citep{luo2011}. 
However, the increment observed in the experiments is smaller than the resolution of the current data (i.e., the 
error bars depicted in the figure). Therefore, we cannot conclusively confirm the experimental findings with the 
available simulation data.

\subsection{Size-based separation}\label{subsec:sizeBasedSep}
We have established that the critical offset $b_c$ increases with increasing particle size, and/or 
decreasing the constriction aperture. Particularly, we have concluded that, for the same constriction 
aperture, two particles of different sizes would exhibit different critical offsets. From the 
perspective of size-based separations, and for the sake of specificity, let $b_{c1}$ and $b_{c2}$ be the 
values of these critical offsets for particles of radii $a_1$ and $a_2$, respectively. If $a_1<a_2$, then 
from the previous discussion, it follows that $b_{c1}<b_{c2}$, assuming the same range of non-hydrodynamic 
interactions $\epsilon$ (such as the amplitude of surface roughness). If these particles are made to move 
past the obstacle with the same initial offset satisfying $b_{c1}<b_{in}<b_{c2}$, the larger particle would 
get displaced onto the critical trajectory and would travel with a final offset $b_{c2}$, while the smaller 
particle would not get affected by the non-hydrodynamic interactions, and continue with the final offset $b_{in}$, 
thus separating the two particles spatially, along the $z$-direction. Another possibility would be an initial 
offset satisfying $b_{in}<b_{c1}<b_{c2}$, which leads to separation since both particles would move along the 
corresponding critical trajectories downstream to the obstacle. The lateral displacement between the two particles 
in this case is $(b_{c2}-b_{c1})$ and is larger than that in the previous case (i.e., $b_{c2}-b_{in}$). Therefore, 
the spatial resolution of separation corresponding to this latter possibility is maximum. As shown in figure 
\ref{fig:bcVsaLB2}, the critical offset is more sensitive to changes in particle radii in the pressure-driven case 
a constant force. Therefore, the spatial resolution would be higher between the two particles when the driving 
field is the fluid flow.

\section{Summary}
In summary, we have studied the behavior of spherical particles of different sizes moving in a channel 
through a constriction between a fixed spherical obstacle and a plane wall. We have investigated cases 
pertaining to different particle-to-obstacle aspect ratios and constriction apertures. We observe that the 
particles reach closer to the obstacle as the particle size increases or the constriction aperture reduces 
in size. Our simulations also show that a particle driven by a constant force in a quiescent fluid reaches 
closer to the obstacle than the one moving with a pressure-driven flow. However, in the 
case of a pressure-drop driven flow, we observe a larger range of change in $b_c$. We have discussed the 
implications of these observations assuming that a hard-wall potential model fairly represents the repulsive 
non-hydrodynamic interactions. We have inferred that particle reaching closer to the obstacle (as a function 
of particle size or the constriction aperture) is equivalent to increasing the critical offset associated 
with the particle. Such dependence of the critical offset on the particle size serves as the basis for 
size-based separation. Finally, the separation resolution is higher in the case of a fluid flow driving the 
particles than a constant force in a quiescent fluid.

\begin{acknowledgements}
We thank Prof. A. J. C. Ladd for making the LB code, {\em Susp3d}, available. This work is 
partially supported by the National Science Foundation Grant Nos. CBET-0731032, CMMI-0748094 and CBET-0954840. 
This work used the resources of the National Energy Research Scientific Computing Center, which is supported by 
the Office of Science of the U.S. Department of Energy under Contract No. DE-AC02-05CH11231.
\end{acknowledgements}

\end{document}